\documentclass{article}

\usepackage{arxiv}

\usepackage[utf8]{inputenc} 
\usepackage[T1]{fontenc}    
\usepackage{hyperref}       
\usepackage{url}            
\usepackage{booktabs}       
\usepackage{amsfonts}       
\usepackage{nicefrac}       
\usepackage{microtype}      
\usepackage{lipsum}		
\usepackage{graphicx}
\usepackage{subfig}
\usepackage{natbib}
\usepackage{doi}

\title{Deep Learning for Reduced Order Modelling and Efficient Temporal Evolution of Fluid Simulations}

\author{ {Pranshu Pant}\\
	Department of Mechanical Engineering\\
	Carnegie Mellon University\\
	Pittsburgh, PA, USA \\
	\texttt{ppant@andrew.cmu.edu} \\
	\And
    {Ruchit Doshi}\\
	Department of Mechanical Engineering\\
	Carnegie Mellon University\\
	Pittsburgh, PA, USA \\
	\texttt{ruchitsd@andrew.cmu.edu} \\
	\And
    {Pranav Bahl (Research Intern)}\\
	Department of Mechanical Engineering\\
	Carnegie Mellon University\\
	Pittsburgh, PA, USA \\
	\texttt{bahlpranav24@gmail.com} \\
	\And
    {Amir Barati Farimani}\\
	Department of Mechanical Engineering\\
	Carnegie Mellon University\\
	Pittsburgh, PA, USA \\
	\texttt{barati@cmu.edu} \\

}

\renewcommand{\shorttitle}{\textit{DL-ROM for Efficient Temporal Evolution of Fluid Simulations} }

\hypersetup{
pdftitle={A template for the arxiv style},
pdfsubject={q-bio.NC, q-bio.QM},
pdfauthor={David S.~Hippocampus, Elias D.~Striatum},
pdfkeywords={First keyword, Second keyword, More},
}

\begin{document}
\maketitle

\begin{abstract}
Reduced Order Modelling (ROM) has been widely used to create lower order, computationally inexpensive representations of higher order dynamical systems. Using these representations, ROMs can efficiently model flow-fields while using significantly lesser parameters. Conventional ROMs accomplish this by linearly projecting higher order manifolds to lower-dimensional space using dimensionality reduction techniques such as Proper Orthogonal Decomposition (POD). In this work, we develop a novel deep learning framework DL-ROM (Deep Learning - Reduced Order Modelling) to create a neural network capable of non-linear projections to reduced order states. We then use the learned reduced state to efficiently predict future time steps of the simulation using 3D Autoencoder and 3D U-Net based architectures. Our model DL-ROM is able to create highly accurate reconstructions from the learned ROM and is thus able to efficiently predict future time steps by temporally traversing in the learned reduced state. All of this is achieved without ground truth supervision or needing to iteratively solve the expensive Navier-Stokes(NS) equations thereby resulting in massive computational savings. To test the effectiveness and performance of our approach, we evaluate our implementation on five different Computational Fluid Dynamics (CFD) datasets using reconstruction performance and computational runtime metrics. DL-ROM can reduce the computational runtimes of iterative solvers by nearly two orders of magnitude while maintaining an acceptable error threshold.
\end{abstract}


\section{Introduction:}
Physics based simulation models are proving to be of paramount importance across various engineering and scientific disciplines. These models have often found their significance in the areas of Aerospace design, HVAC, Cardiovascular flows, Electronics, Turbo-machinery etc ~\cite{Intro_1}. The necessity to make an explicit engineering/scientific decision involving complex design processes, empirical discoveries, experimental design etc., requires the resolution of the predictions to be very high. These high-fidelity predictions demand high temporal and spatial resolutions which often leads to modelling of various complex non-linear processes into a very large scale dynamical model. The simulation of such models is often associated with full-order-models (FOM) which are based on parametrized system of physics governing partial differential equations (PDE). 
\\
Whenever high-dimensional FOM find their applications pertaining to real time or multi-query scenarios, there is an overwhelming increase in computational cost/burden which imposes restriction on expeditious simulation result generation ~\cite{fresca_Milano,Lee_Sandia}. Some common examples pertaining to these scenarios are uncertainty quantification ~\cite{part2_ref4,part2_ref5}, flow control~\cite{part2_ref1,part2_ref2,part2_ref3}, multi-fidelity optimization techniques etc. Since conventional FOM gives rise to increased utilization of the computational resources, it is often prohibitive to use such models in various areas of study. Hence, there is an inherent need for devising methodologies that overcome this issue and present a reduced representation of high order dynamical system to a lower dimension. The objective of the reduced representation is to project the physical features of a system comparable to a FOM, with minimum loss of information to a lower dimensional space/manifold. The approach is therefore referred to as reduced order modelling and is often abbreviated as (ROM). There have been various attempts to model such approaches in the past  ~\cite{part1_ref1,part1_ref2,part1_ref3,part1_ref4,part1_ref5,part1_ref6}. The development of reduced order models is a challenging task, owing to the fact that models are often not robust enough to handle parameter alterations and are neither cost-effective when it comes to dealing with complex time varying physical phenomenons. One of the popular approaches is to express the system as a linear amalgam of the basis functions formed with the help of snapshots (series of temporal results generated corresponding to a parameter space) generated using high-fidelity expensive simulations involving FOMs. These approaches comes under the category of Projection-based-ROMs wherein the focus is to achieve a transformed space formed from the reduction of high DOF (degrees of freedom) of the physics governing PDEs ~\cite{san_osu}. The reduction results in the computation of a low dimensional trial subspace corresponding to the state of the system. The computation of low-dimensional subspace is inexpensive when approximating solutions which are projected w.r.t different points in the parameter space.  ~\cite{Lee_Sandia}, this is achieved while imposing high dimensional FOM residuals orthogonal to a lower dimensional space:\emph{ Linear test subspace }. However, projection based ROMs are predominantly linear, this is because they project onto a linear subspace, as discussed above, to extract the reduced representation of high order dynamics. 
\\ 
\begin{figure*}[t]
  \centerline{\includegraphics[width=0.7\textwidth]{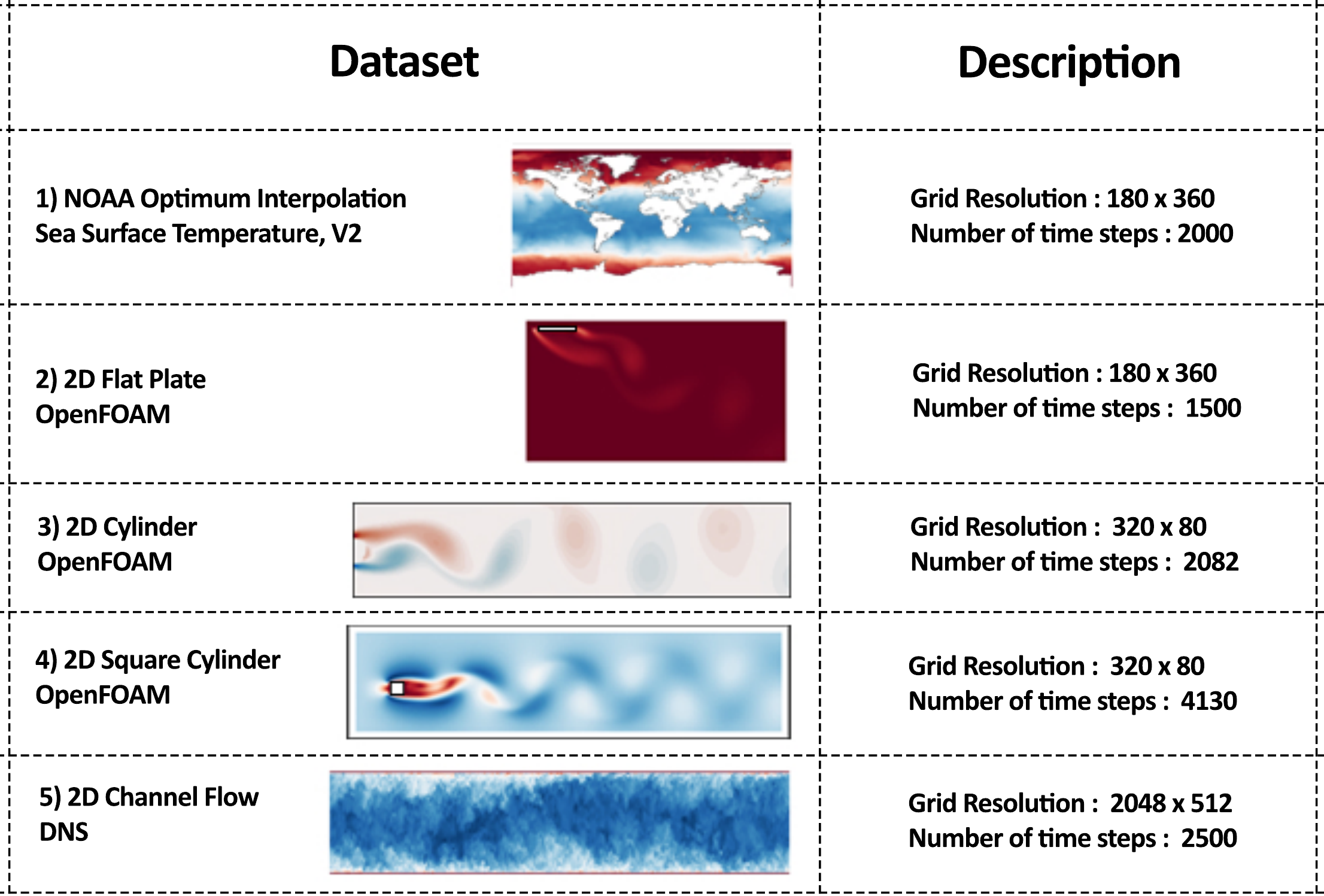}}
  \caption{Datasets with different simulation conditions, 1) NOAA - SST - Weekly Mean Sea Surface Temperature, 2) Flow over plate - with vortex shedding, 3) 2D Cylinder - with Von Karman vortex, 4) 2D Square Cylinder - with vortex shedding, 5) Channel Flow - Turbulent Flow along a channel, OpenFOAM - Custom Designed dataset with Von Karman vortex, studied to evaluate the Deep Learning Based approach for Reduced Order Modelling. These datasets are 2-Dimensional and contain u velocity data (X-direction).}
\label{fig:dataset}
\end{figure*}

Among numerous projection based ROMs, Proper orthogonal decomposition (POD) ~\cite{part3_ref1,part3_ref2,part3_ref3,part3_ref4} has found its acceptance among many academicians. It is analogous to its counterparts PCA (Principal component analysis)~\cite{hotel}, empirical orthogonal functions~\cite{loren} and Karhunen-Loeve expansion~\cite{loev}. The eigen-decomposition of the snapshot matrix is carried out using singular value decomposition (SVD) technique under the regime of this ROM \cite{taira2017modal}. The result of the following is a Linear ROM, whose predictions are computed in linear trial subspace and hence, are generated by linear super-imposition of the POD modes. One of the similar approaches to POD is DMD (Dynamic mode decomposition)~\cite{DMDER,DMD2} wherein the objective is the same, to represent the generated high order data to a low dimensional subspace but have well defined dynamics corresponding to the subspace. DMD represents significant features of both, POD and Discrete Fourier Transforms (DFT) and is often very successful in extracting physical insights of the system in the form spatio-temporal coherent structures\cite{rowley,taira2017modal}. DMD although having a wide array of applications ranging from robotics\cite{berger2014dynamic,berger2015estimation}, neuroscience\cite{brunton2016extracting}, epidemiology\cite{bistrian2019processing} to image-video processing\cite{schmid2011applications}, is not very helpful in application pertaining to flow control, state prediction, estimation etc. There are various hybrid approaches associated with POD such as POD-GP~\cite{PODGP1,PODGP2,PODGP3,PODGP4}, wherein the Galerkin-Projection technique is used for the truncation of high-order discretizations of Partial differential equations to a reduced state which is also governed by a set of ordinary differential equations to generate temporal coefficients. This methodology makes use of POD modes which is an orthogonal approach for projection, hence the loss of information inevitable due to the linearity of the approach. Non-orthogonal approach have also been introduced in the past, for example the Petrov-Galerkin~\cite{petrov,taira} model wherein the property of bi-orthogonality is used to obtain the ROM. Among other approaches, there is a popular approach known as Koopman Operator theory~\cite{koop,taira}. This theory has found its application across various complex non-linear systems, in its early efforts from characterization of the dynamics of Hamiltonian functions\cite{koopman1931hamiltonian} to an infinite-dimensional linear operator to the decomposition of dynamics pertaining to fluids. The approach of this method is to find a subspace where the high order dynamics can be linearized and uncoupled\cite{taira2017modal}.  
\begin{figure*}[t]
  \centerline{\includegraphics[width=1\textwidth]{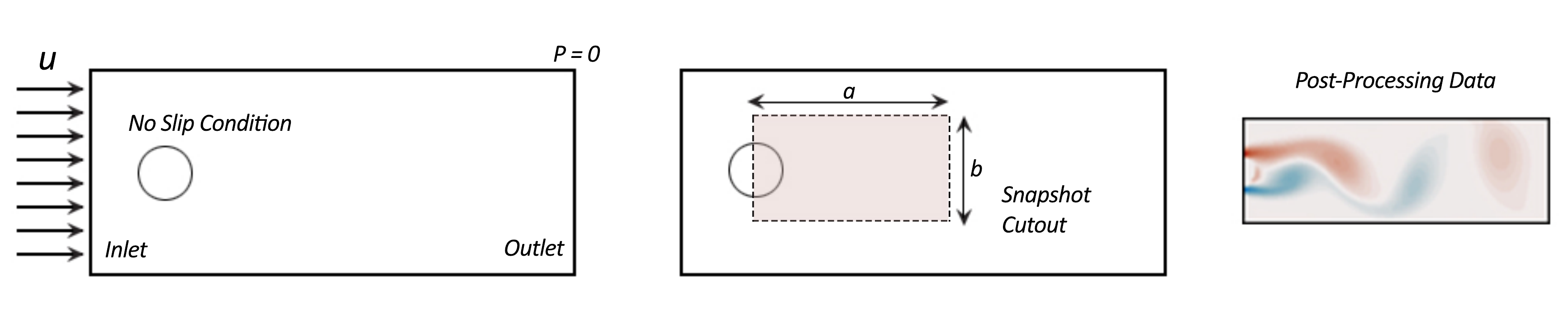}}
  \caption{Schematic representation of the computational domain with boundary conditions at the inlet and the outlet. The bluff body has a characteristic length and a no-slip boundary is considered at the wall of the bluff body. Zero pressure gradient at the inlet and a zero velocity gradient at the outlet are considered for the computational study. Schematic representation of the snapshot cutout $a \times b$ has also been represented wherein the values of $a$ and $b$ are different corresponding to the features exhibited by different computational studies.}
\label{fig:comp_domain}
\end{figure*}

More recently, deep learning has also been used for developing ROMs \citep{WU2020112766, wang2018model, mohan2018deep, SAN2018681, murata_fukami_fukagata_2020}. Due to their inherent non-linear formulation, deep neural networks are highly proficient at compression tasks with dimensionality reduction of fluid simulations being one such task. Thus, neural networks particularly autoencoders can perform similarly to linear projection methods like POD and DMD by projecting the FOMs to reduced order states. Moreover, due to use of non-linear activation functions, neural networks can perform non-linear projections and can thus be used to create superior reduced order representations.Through this work, we wish to utilize the recent advancements made in the realm of deep neural networks and apply it to the field of ROMs and computational fluid dynamics (CFD) in general. We aim to create a non-linear dimensionality reduction neural network that can extract high quality, reduced order embeddings of the complex fluid-flow phenomenon. We can then use these embeddings to create high-fidelity reconstructions of fluid data. Furthermore, we can also use these embeddings to temporally traverse in the reduced state and yield the full order reconstructions at future steps without solving the computationally expensive Navier-Stokes(NS) equations. The mathematical representation of our model has been presented below. The predicted output at time step $t+1$ can be represented as $\hat{y}^{t+1}$ wherein the $y$ is the state vector. The deep learning model and its parameters i.e. its weights and biases are represented as $\varphi(;\theta_\varphi)$. The input to the DL model is concatenated form of state vector at previous time-steps from $t-k^{th}$ time-step to the $t^{th}$ time-step where $k$ represents the number of previous snapshots taken, and can be represented as $\bigodot_{i=0}^k y_{t-i}$. The error term represents the difference between the predicted and original state vector.

\begin{equation}
 y^{\ t+1} = \varphi(\bigodot_{i=0}^k y_{t-i};\theta_\varphi) + e
\end{equation}
\vspace{3mm}
\begin{equation}
error =  y^{\ t+1} - \hat{y}^{\ t+1}
\end{equation}
\vspace{-7mm}

\vspace{-3mm}
\section{Datasets}
\begin{figure*}[t]
  \centerline{\includegraphics[width=0.8\textwidth]{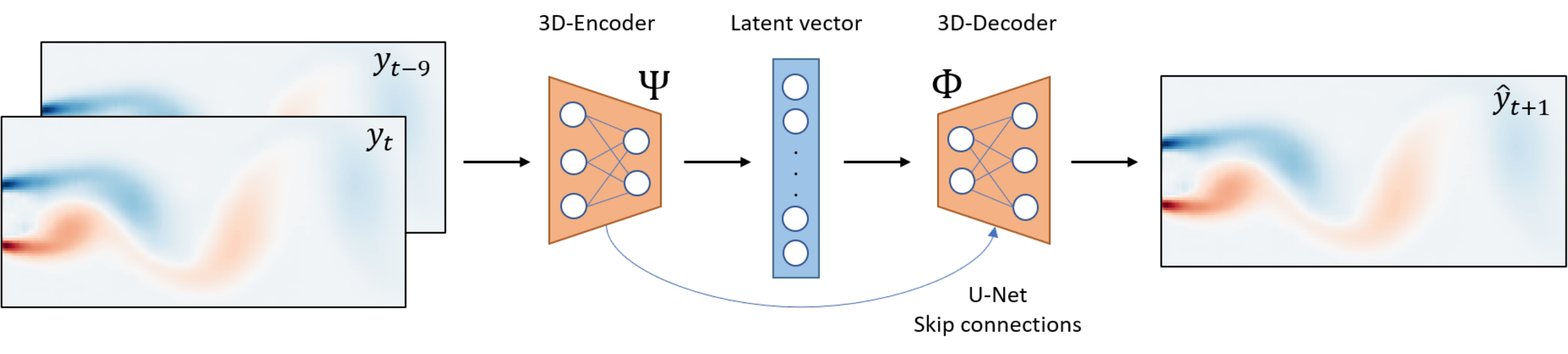}}
  \caption{Framework for the transient Reduced order Model (DL-ROM). 10 snapshots ( [$y_{t-9}$, $y_t$]) of the previously solved CFD data are stacked and used as input to the model. The model then uses a 3D encoder architecture $\Psi $ to reduce the high-dimensional CFD data to reduced order latent vector. This latent space is then deconvolved using a 3D-decoder $\Theta $ to produce the higher order CFD prediction at timestep t+1 ($\hat{y}_{t+1}$). }
\label{fig:framework_nn}
\end{figure*}

To assess the capabilities of our novel deep learning approach for producing a reduced order model and for utilizing the learned reduced state to predict future time steps of a simulation accurately and efficiently, we present a study of five datasets with different simulation conditions. These datasets have been studied well in the literature, making them well suited for benchmarking this application. The processed datasets and code used for this study are available on GitHub at \href {https://github.com/pranshupant/DL-ROM}{github.com/pranshupant/DL-ROM}. More details about these datasets are outlined below.


\subsection{Numerical Experiments: OpenFOAM}
\subsubsection{Flow around 2D Circular Cylinder}


For our first example, we consider a bi-dimensional flow past a circular cylinder at Reynolds' number $Re = 100$. The simulation is a well known canonical problem and the characteristics of the problem are periodic vortex shedding behind the bluff body. The simulation was done using a Reynold's Average (RANS) approach. Incompressible Navier-Stokes Equations were solved for laminar flow around the cylinder using the OpenFOAM \cite{OpenFOAM} solver icoFoam. The solver uses a Pressure-Implicit and Splitting-operators algorithm to solve the momentum and continuity equation.

\begin{equation}
        \nabla.\textup{u} = 0
\end{equation}
 \vspace{0mm}
\begin{equation}
       \frac{\partial}{\partial t} \textup{(u)} + \nabla.\textup{(u}\otimes\textup{u)} - \nabla.\textup{(\emph{v}}\nabla\textup{u)} = -\nabla.\textup{p}
\end{equation}

The characterstic length/diametre of the cylinder is 1 metre. The computational domain is divided into a fine Hex-mesh of 63,420 elements, which was generated using the blockMesh utility in OpenFOAM .The inlet is at a distance of 8 units from the centre of the cylinder and the outlet is at 25 units. The kinematic viscosity of the fluid is 0.01 and an uniform inlet velocity of 1 unit in the positive X-direction. There is an ambient pressure condition at the outlet and a no-slip boundary condition on the cylinder. The time-step for the simulation is kept at $\delta t = 0.008$ seconds. The simulation was carried out parallely over 16 threads and took over 0.12 seconds per iteration. After the simulation reaches the stage of periodicity, the relevant features of the system are extracted by cutting out a snapshot of 2 units x 8 units from the whole domain, encompassing only the vortices. The snapshot is further discretized into 80 x 320 linearly interpolated points in the y and x direction. The objective here is to recover the Z-vorticity field from the temporal information gathered from previous similar snapshots. A total of 2082 snapshots were generated for the training of the model with the temporal resolution of $\delta t$.           


\subsubsection{Flow around 2D Square Cylinder}

For the second example, we consider a two-dimensional flow past a square cylinder at Reynolds' number $Re = 100$. 
The simulation problem presents the same physical characteristics as the flow past circular cylinder wherein there is a vortex shedding phenomenon observed behind the bluff-body. The simulation setup was solved using a Reynold's Average (RANS) approach. Same as in the previous case, Incompressible Navier-Stokes Equations were solved for laminar flow around the square cylinder using the OpenFOAM \cite{OpenFOAM} solver $icoFoam$. The governing equations of the system can be referred from section $1.$ $Flow$ $around$ $2D$ $Circular$ $Cylinder$ wherein the continuity and momentum equation are specified. The characterstic length here for the square cylinder is 0.5 metre. The computational domain is divided into a fine Hex-mesh of 73,750 elements, which was generated using the $blockMesh$ utility in OpenFOAM .The inlet is at a distance of 2.5 units from the centre of the square cylinder and the outlet is at 16.5 units. The kinematic viscosity of the fluid is 0.02 and an uniform inlet velocity of 1 unit in the positive X-direction. The boundary conditions are, ambient pressure condition at the outlet and a no-slip boundary condition on the square cylinder. The time-step for the simulation is kept at $\delta t = 0.01$ seconds. The simulation was carried out parallel over 16 threads, which took over 0.25 seconds per iteration. After the simulation reaches the stage of periodicity, the relevant features of the system are extracted by cutting out a snapshot of 2 units x 8 units from the whole domain encompassing the relevant fluctuations in the velocity field. The snapshot is divided linearly into 80 x 320 interpolated points in the y and the x direction. The objective here is to recover the velocity field from the temporal information gathered from previous 9 snapshots. A total of 4087 snapshots were generated for the training of the model with the temporal resolution of $\delta t$.  

\subsubsection{Flow over 2D Plate}

For third example, we consider a two-dimensional flow over a 2D Plate.
The simulation problem put forwards the physical characteristic of an inclined plane wherein the vortex structure formation can be observed behind the body. The simulation setup was solved using a Reynold's Average (RANS) approach wherein the K-epsilon turbulence model was used to model the turbulence. Incompressible Navier-Stokes Equations were solved for flow over the flat plate using the OpenFOAM \cite{OpenFOAM} solver $pimpleFoam$. The turbulence model used here is a two transport equation, Linear-eddy-viscosity closure model wherein the equations used are: 1) Turbulent kinetic energy equation\cite{Kepsilon} - $k$ , 2) Turbulent kinetic energy dissipation rate equation\cite{Kepsilon} - $\epsilon$ and 3) Turbulent viscosity equation\cite{Kepsilon} - $\nu_t$.

\begin{equation}
    \frac{D}{Dt} (\rho k) = \nabla \cdot(\rho D_k \nabla k) + P - {\rho\epsilon}
\end{equation}
 \vspace{0 mm}
\begin{equation}
      \emph{v}_t  = C_\mu \frac{k^2}{\epsilon}
\end{equation}
 \vspace{0 mm}
\begin{equation}
      \frac{D}{Dt} (\rho \epsilon) = \nabla \cdot(\rho D_\epsilon \nabla \epsilon) + \frac{C_1 \epsilon}{k} (P + C_3 \frac{2}{3} k\nabla \cdot \textup{u} - C_2  \rho \frac{\epsilon^2}{k}) 
\end{equation}\\

The characteristic length here for the 2D plate is 1 metre and the width of the plate is 0.04 metre. The computational domain is divided into a fine Hex-mesh of 70,600 elements, which was generated using the $blockMesh$ utility in $OpenFOAM$ .The inlet is at a distance of 2 units from the 2D Plate and the outlet is at the distance of 8 units. The kinematic viscosity of the fluid is 0.00005 and an uniform inlet velocity of magnitude 1 unit and the direction of the velocity is 45 degrees to the positive X-direction. There is an ambient pressure condition at the outlet boundary and a no-slip boundary condition on the 2D plate. The time-step for the simulation is kept at $\delta t = 0.005$ seconds. The simulation was carried out parallel over 16 threads which resulted in 0.128 seconds per iteration. After the simulation reaches the stage of periodicity, the relevant features of the system are extracted by cutting out a snapshot of 4 units x 8 units from the computational domain encompassing only the vortex structures. The snapshot is further divided equally into 180 x 360 linearly interpolated points in the y and the x direction respectively . The objective here is to recover the vorticity magnitude from the temporal information gathered from previous time step's snapshots. A total of 1500 snapshots were generated for the training of the model with the temporal resolution of $\delta t$.  
 \begin{figure*}[t]
  \centerline{\includegraphics[width=1.1\textwidth]{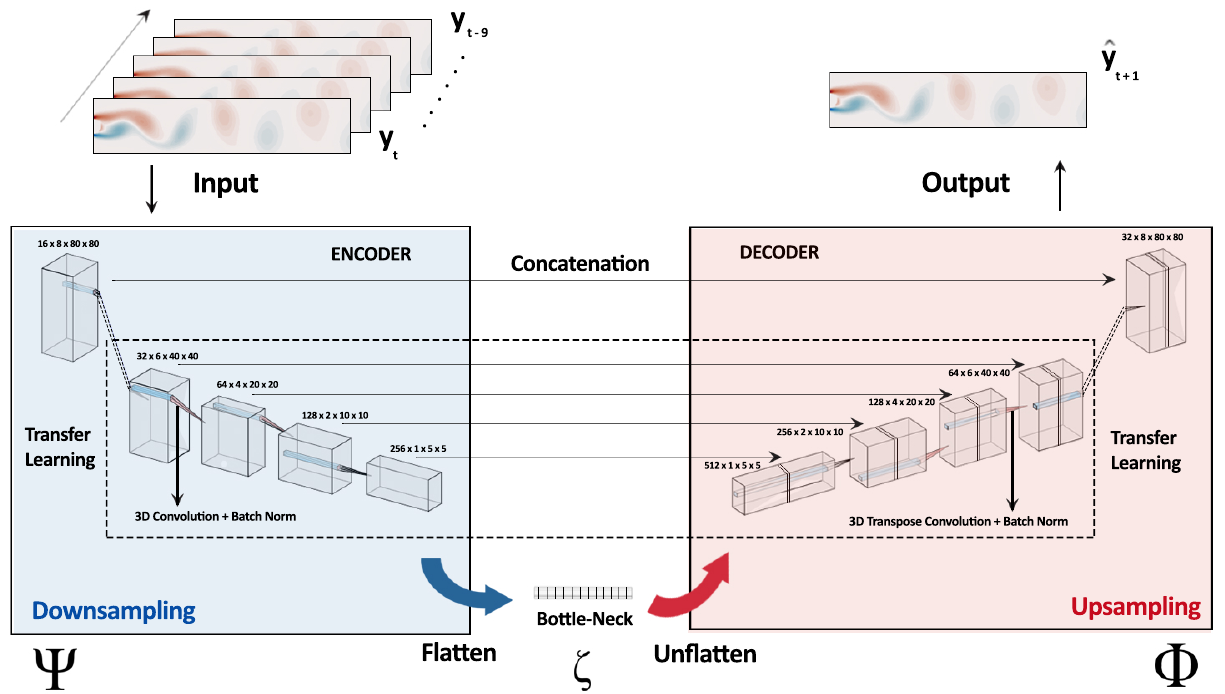}}
  \caption{3D Autoencoder based UNet Model Architecture for our framework DL-ROM. 10 timesteps are concatenated to generate temporal context as the input to the  architecture. Each block represents the intermediate size of the data. The arrows represents the skip connections between encoder and decoder part of the architecture. The bottleneck represents a 1D vector of the Reduced Order States of the input.}
\label{fig:architecture}
\end{figure*}
              


                    
\subsection{2D Channel Flow dataset}

This dataset consists of a 2D Direct Numerical Solutions (DNS) of channel flow in a grid size of 2048 x 512. This DNS dataset was obtained from the Johns Hopkins Turbulence Database (JHTDB) \citep{doi:10.1080/14685248.2015.1088656, 10.1145/1362622.1362654,li2008public}. The simulation assumes the fluid used to be incompressible. The Navier-Stokes equations are solved using the Fourier-Galerkin method and 7th-order B-spline collocation method in the x-z and (y) direction respectively. The simulation has bulk velocity=1 and imposes a pressure gradient of 0.0025. Once the simulation reaches a statistically stationary state, 3 component velocity points are added to the database for recording the dataset. The frames are stored at every 5 time-steps of the DNS which corresponds to about one channel flow-through time. To make this DNS dataset amenable for use with our deep learning model we sample the dataset on a uniform grid of 512 x 128 at 2500 timesteps.



\subsection{NOAA Optimum Interpolation (OI) SST, V2}

The NOAA (OI) Sea Surface Temperature V2 data-set has been made publicly available by the Physical Sciences Division at NOAA. The temporal resolutions available for the data-set are weekly, monthly and monthly long-term mean data. The weekly-data is centered on Wednesday for the brief period of 1990-2011 and on Sunday for the period of 1981-1989. For this study, the temporal resolution of the data-set was chosen as ~7 days i.e. weekly data from 1981-2011. The biases of the satellite are tuned as directed in the literature ~\cite{OISST_1988,OISST_1993} following which the data-set is generated through analysis of in-situ and satellite observations. The dataset contains Sea Surface Temperature data in the units of deg Celsius over the globe, the uncertainty with reference to the real world data can be expected in the dataset.  The spatial resolution of the data is 1-deg latitude and 1-deg longitude leading to 180 x 360 grid points. 2000 snapshots of this data are utilized to create our dataset that is used for training the neural network.

\section{Methodology}
\subsection{Existing Approach}
Existing Deep Learning approaches for reduce order modeling and reconstruction of the same time-step of the high order fluid flow utilize autoencoders \citep{8616075}.   
To predict future time steps of a simulation from these reduced order states, Convolutional LSTM Autoencoders \citep{maulik21} are used. These models are a amalgamation of three separate models which are combined to achieve the desired reduce ordered states and predict the future timesteps in the same input dimensional space. \\

\vspace{-5mm}
\subsubsection{Autoencoders}
The main objective of the autoencoder is to perform dimensionality reduction on the FOM CFD simulations using the large dataset and capture the reduced states of the current timestep. The dimensionality reduction, in general, can be a loss compression process, for instance popular methods such as Proper Orthogonal Decomposition (POD) can represent the dominant energetic dynamics in the first few eigen vectors. These eigen values can be represented as the reduced order states and used in further analysis. However, POD projects the datasets into a linear manifold, whereas autoencoders use non-linear activations which are non-linear maps having very high compression ratio. The architecture of the Autoencoders can be designed using different feature extracting layers such as fully-connected (FCN) and convolutional neural networks (CNN). Depending upon the type of data and its input format, the autoencoder architecture can be varied. The autoencoder architecture can be subdivided into two major parts namely the encoder and the decoder. Two popular variants of autoencoders are as follows. \\

\begin{figure*}[t]
  \centerline{\includegraphics[width=0.8\textwidth]{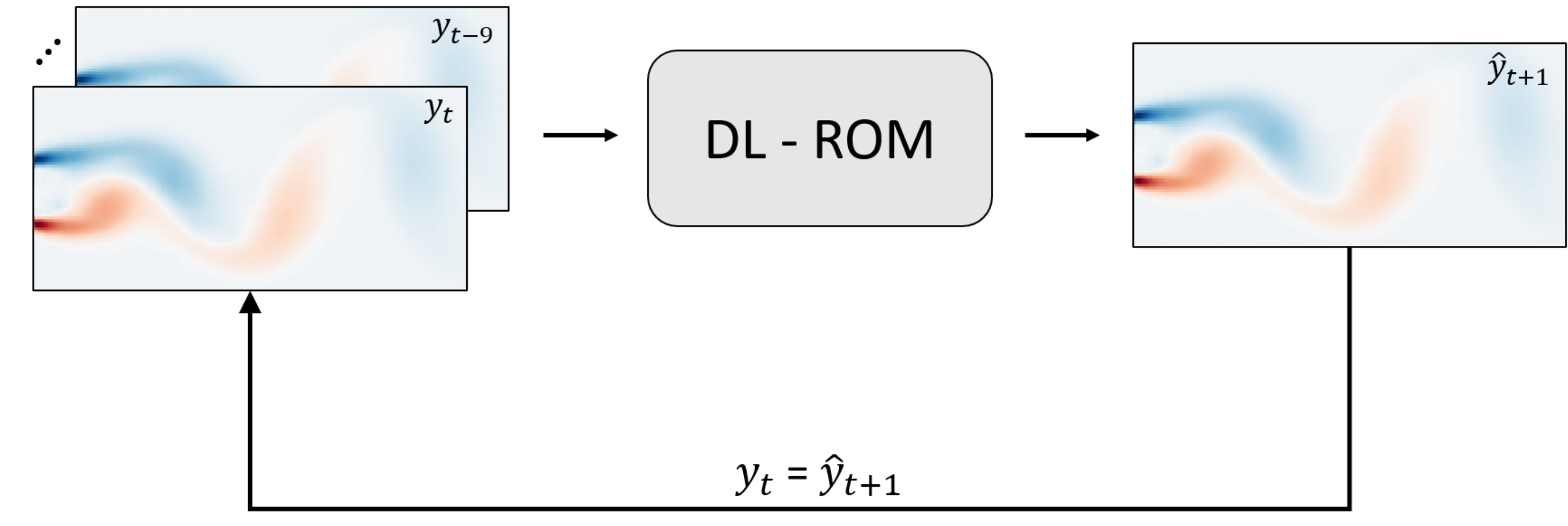}}
  \caption{DL-ROM can be used to for in the loop prediction of future simulation timesteps. If the model is given the result of the first 10 timesteps, it can progressively predict the future timesteps by appending the result of time (t+1) into the original 10 timesteps}
\label{fig:rec_dlrom}
\end{figure*}

             
\subsubsection{Multi-Layered Perceptron Autoencoder:} 
Here, the architecture of the encoder is a series of fully connected layers wherein the input is a one-dimensional vector representing the fluid flow. The middle layer of the architecture, also known as the bottleneck layer, represents the compressed state of the high order dynamical systems. These compressed states represent the reduced ordered states of the input, compressed on a non-linear manifold. The decoder takes these reduced states from the bottleneck layer and reconstructs the input in the original dimensional space using another network of fully-connected layers that is also referred to as a decoder. \\

                    
\subsubsection{Convolutional Autoencoders:} 
The Multi-Layered Perceptron Autoencoder can be useful if the input is a one-dimensional vector and does require to account for the local and spatial characteristics. However, if the input dimensional space is 2D or higher, to account for local features, convolutional layers can be very useful because of its space-invariant properties. Also, the weight sharing property of CNNs make them computationally efficient in comparison to their fully connected layers counterparts.The architecture of Convolutional Autoencoders \citep{inproceedings} is very similar in its approach to the Multi-Layered Perceptron Autoencoder, replacing each fully connected layer with convolutional layers. The features extracted by a series of convolutional layers in the encoder are then transformed into a bottleneck layer using one fully-connected layer. The decoder uses transpose convolutions to upsample the bottleneck layer to the original higher dimensional space.    

\subsubsection{Long Short-Term Memory networks}
To utilize the latent vector or the reduced states of the input and learn the future latent predictions, we need to preserve the sequential information contained in the transient CFD simulations. Thus the goal of predicting the evolution of a high order dynamical system from its reduced order states becomes one of sequence modelling in deep learning. Recurrent Neural Networks (RNN) \citep{Sherstinsky_2020} is class of artificial neural networks where connections between nodes form a directed acyclical graph along the temporal dimension. This temporal information is thus, added as a new dimension in RNNs to solve the sequence modelling problem. The Long Short-Term Memory (LSTM) neural network \cite{HochSchm97} is a special variant of the RNN, which improves network performance by addressing some drawbacks of the vanilla RNNs. The RNNs suffer from bottleneck instabilities like vanishing gradients and have less memory retention properties.The LSTM network contains LSTM cells which are made by three gates namely - input gate, output gate, and the forget gate to regulate the flow of information. A separate memory cell is maintained in LSTM cells which helps in longer memory retention. The input gate does selectively filtering of new information, the forget gate removes the redundant information and the output gate adds the essential information to the next cell. This arrangement of gates and selective information control is also the key reason why LSTMs do not suffer from the vanishing gradient problem which plagues traditional RNNs. As a result, LSTMs are a powerful tool to model sequential datasets including application in transient fluid simulations \citep{maulik21}.

\begin{equation}
    \mathbf c_t = \mathbf F_t\odot \mathbf c_{t-1}+ \mathbf I_t\odot \mathbf{G}_t
\end{equation}
 \vspace{0 mm}
\begin{equation}
     \mathbf{G}_t = \tanh (\mathbf \chi_t \Gamma_{x c} + \mathbf k_{t-1} \mathbf W_{h c} + \mathbf b_c)
\end{equation}
 \vspace{0 mm}
\begin{equation}
     \mathbf I_t = \sigma( \mathbf \chi_t \Gamma_{x i} + \mathbf k_{t-1} \mathbf W_{hi} + \mathbf b_i)
\end{equation}
 \vspace{0 mm}
\begin{equation} 
     \mathbf F_t = \sigma(\mathbf \chi_t \Gamma_{x f} + \mathbf k_{t-1} \mathbf W_{hf} + \mathbf b),
\end{equation}
 \vspace{0 mm}
\begin{equation}
     \mathbf o_t = \sigma \left( \mathbf \chi_t \Gamma_{x o} + \mathbf k_{t-1} \mathbf W_{ho} + \mathbf b_o \right)
\end{equation}
 \vspace{0 mm}
\begin{equation} 
     \mathbf k_t = \mathbf o_t \odot \tanh (\mathbf c_t),
\end{equation} \\

\begin{figure}[t]
  \centerline{\includegraphics[width=3.8 in]{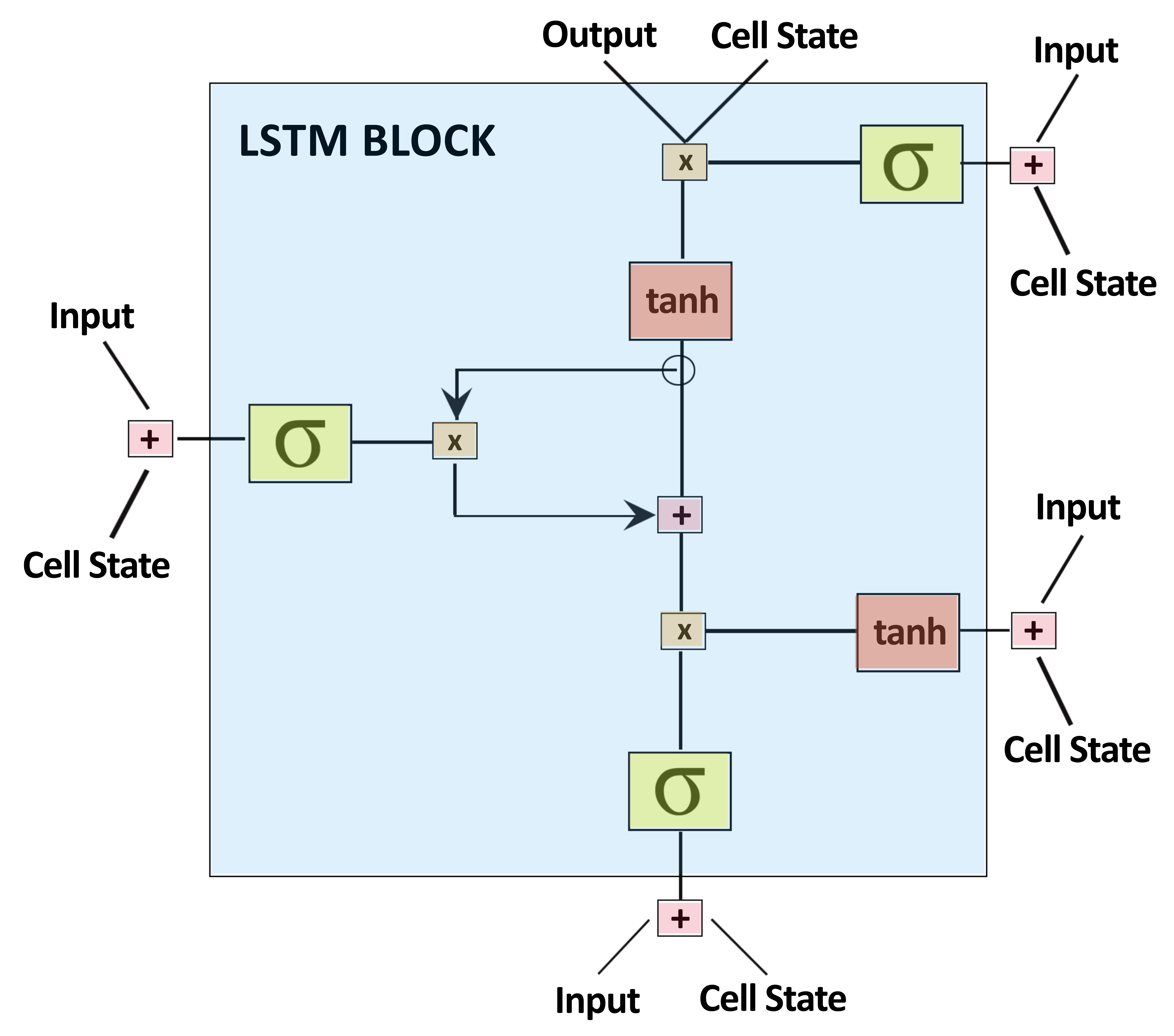}}
  \caption{Schematic representation of a typical LSTM block. Output of the previous block and the cell state are concatenated for the input to the Input, Forget and Output gate of the block. Update in the candidate cell state is carried out in the form of addition, thereby preserving information for long-term without facing vanishing gradient.}
\label{fig:LSTM}
\end{figure}

\subsection{Our Approach}

\begin{figure*}[t]
 \centering
    \includegraphics[width=0.95\textwidth]{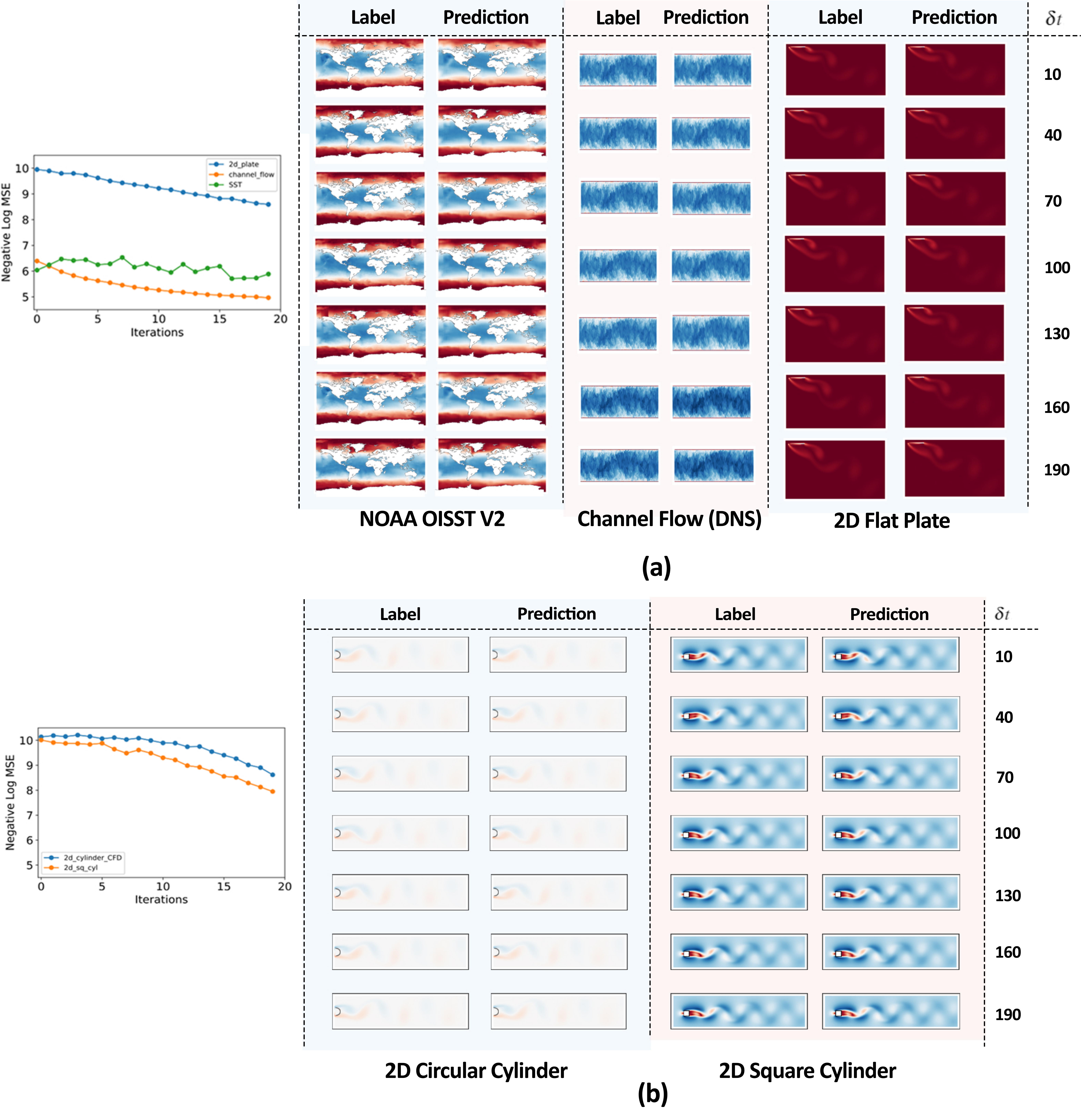}
    \caption{Results obtained on the 5 datasets using our Deep Learning based approach for Reduced Order Modelling. Each dataset is split into training and validation subsets. The labels and the corresponding predictions presented are from the validation split which are not used for training. Progression of MSE with timesteps evaluated on the validation dataset. The DL-ROM model is provided with data for only the initial timestep. The simulation evolves using the predictions of the previous timesteps without supervision from the ground truth values. As expected the value of negative log MSE gradually decreases over time due to accumulation of errors. Note that decreasing lineplots of Negative Log MSE represent increasing MSE values.}
    \label{fig:domain}
\end{figure*}

Even though the existing approaches described in the previous section could be used to achieve a deep learning framework that is similar to the one we envisioned by us these approaches are plagued by certained disadvantages that we tackle with our proposed approach.
The use of LSTMs for temporal sequences can have many disadvantages. LSTMs usually take a longer time to train, are computationally expensive and are often difficult to train. Apart from computational complexities, these models tend to overfit on the training data and often suffer from a problem known as exploding gradients. Generally, the fully connected LSTM networks, which take in vectorized reduced order features as input (as described above) are used to learn temporal features. This results in loss of spatial correlation information during the recurrence. To tackle this issue, our approach uses 3D convolutions \cite{hou2017endtoend} which can extract features in both, spatial and temporal axes.\\

\subsubsection{3D Convolutional Autoencoders}

A 3D Convolution is a type of convolution where the kernel slides in 3 dimensions as opposed to 2 dimensions with 2D convolutions. It is mostly used with 3D image/video data that are 4 dimensional with the 4th dimension representing the number of channels. Some use cases for such data are - MRI scans where relationship between a stack of images is to be understood; and a low-level feature extractor for spatio-temporal data like videos for Gesture Recognition, Weather forecast etc. Since, high order dynamic simulations can be assumed as a stack of images, we need to extract spatio-temporal characteristics for producing reduced ordered states and to predict the future time steps. Instead of using LSTMs, which are difficult to train, in this novel approach we make use of 3D Convolutional Auto-Encoders to efficiently predict ROMs and future time steps. Our model architecture also comprises of an encoder and a decoder network. The encoder is made of 5 layers of 3D convolutions, each followed by BatchNorm and ReLu activations. For our convolutional layers we make use of depth-wise separable convolutions \citep{chollet2017xception, 43022}. Depth-wise separable convolutions are similar to conventional CNNs but with one stark difference. Such convolutions use only one filter for every input channel, thereby significantly reduce the number parameters that have to be learned while maintaining similar performance in terms of feature extraction. Next, the strides of the convolutions are kept greater than one to reduce the image resolution (compression). The spatio-temporal features extracted from the encoder are flattened and converted into a one-dimensional feature vector which represents the reduced states. These reduced states are then reshaped and passed onto the decoder to reconstruct the high dimensional future time step of the simulation. The decoder consists of 5 layers of 3D transposed convolutions, each followed by BatchNorm and ReLu activation. Transposed convolutions (deconvolutions) with stride greater than one are used to increase the size of the input data and match it with the original dimensional space. The output of the decoder is a 2D dimensional image which represents a future time step in the simulation. The mathematical foundation of the approach has been presented below. The $\phi(;\theta_\phi)$ is the representation of the convolved decoder network that has the function of projecting the non-linearly learnt reduced state back to the original high-order dimensional space, which in this case would be the predicted state vector at time step $t+1$. The term $\theta_\phi$ represents the decoder's weights and biases i.e. its parameters. The convolved encoder model can be presented as $\psi(w_t;\theta_\psi)$ which is responsible for the compression of the concatenated input of previous time-steps $\bigodot_{i=0}^k y_{t-i}$ to a non-linear reduced state $\zeta$. $\theta_\psi$ again represents the weights and biases i.e. the parameters of the convolved encoder model. $\hat{y}^{\ t+1}$ is the predicted state vector and ${y}^{t+1}$ represents the original state vector at time-step $t+1$.       
\\
\begin{equation}
      	\zeta = \psi(\bigodot_{i=0}^k y_{t-i};\theta_\psi) ;  \phi( \zeta) = \hat{y}^{\ t+1}
\end{equation}
\vspace{3mm}
\begin{equation} 
          \hat{y}^{\ t+1} = \phi(\psi(\bigodot_{i=0}^k y_{t-i};\theta_\psi);\theta_\phi)
\end{equation}

\subsubsection{3D U-Net}
To augment the performance of the 3D Convolutional Autoencoder, we make use of a U-Net architecture \citep{ronneberger2015unet} as shown in Fig.\ref{fig:architecture}. This architecture contains multiple links between its encoder and decoder at each step, wherein the image resolution is same. These links consist of saving snapshots of the weights during the first phase of the network and copying them to the second phase of the network. This makes the network combining features from different spatial regions of the image and allows it to localize the regions of interests more precisely. In our approach, we concatenate 10 frames of the higher order fluid flow taken each after every 10 timesteps of the simulation to account for the temporal context. So, the first input to our model is 0th, 10th, 20th, … up-till 90th frame concatenated together. The target timestep would be the 100th timestep. As consecutive timesteps of the simulations do not result in significant changes in the fluid flow quantities we keep stack frames in steps of 10. Additionally, we fix the depth or temporal context of size 10 (10 frames stacked together) for our approach. Similar preprocessing is done on other inputs. These inputs are then passed on to the model to capture the reduced states and predict the future timesteps. The output of the decoder is then used to compute the loss/error between the target image and model prediction. The loss function used my our model (DL-ROM) is Mean Absolute Error (MAE).The mean absolute error (MAE) of an estimator (of a procedure for estimating an unobserved quantity) measures the absolute average of the of the errors i.e. the average absolute difference between the estimated values and the ground truth.\\ \\
\begin{equation}
   L_{MAE} =  \frac{1}{N} \sum_{n=1}^{N}     \left\| e \right\|_1 
\end{equation}
\vspace{3mm}
\begin{equation}
    NN_{DL-ROM} = min \frac{1}{N} \sum_{n=1}^{N}
          \left\| w^{t+1} - \phi(\psi(\bigodot_{i=0}^k w_{t-i})) \right\|_1 
\end{equation}\\

$NN_{DL-ROM}$ represents the objective function of our DL-ROM framework. From the equation above we can see that it aims to minimize the MAE loss between the prediction and the ground truth for the fluid simulation at time t+1 given inputs for timesteps (t-k, t] (where k = 10). This objective is minimized over N training examples/snapshots of CFD simulation dataset. Finally, to showcase the application of DL-ROM to real world CFD problems we run a looped prediction simulation for 20 timesteps/iterations beyond the final output of the CFD solver Fig\ref{fig:rec_dlrom}. The simulation evolves using the predictions of the previous timesteps without supervision from the ground truth values. In this simulation we input into DL-ROM the last 10 frames of the CFD simulation and subsequently predict the output at the next timestep (t+1). This new prediction is then concatenated with the last 9 CFD timesteps and then sent as input to the model which now predicts the output for timestep (t+2). This process is looped for 20 future iterations (t+20) and the trend of MSE between the ground truth and the looped predictions is subsequently evaluated. This is used to evaluate the performance of our network compared to the CFD ground truth in terms of the solution accuracy and computational runtime efficiency. For training the neural network the Adam optimizer \citep{kingma2014adam} was used due of its faster convergence capacity. A learning rate scheduler was also implemented to make use of smaller learning rates as the number of iteration increased and performance stagnated. The training was initialized with a learning rate of 0.05. 
\begin{figure}
     \centering
     \subfloat[][]{\includegraphics[width=0.5\linewidth]{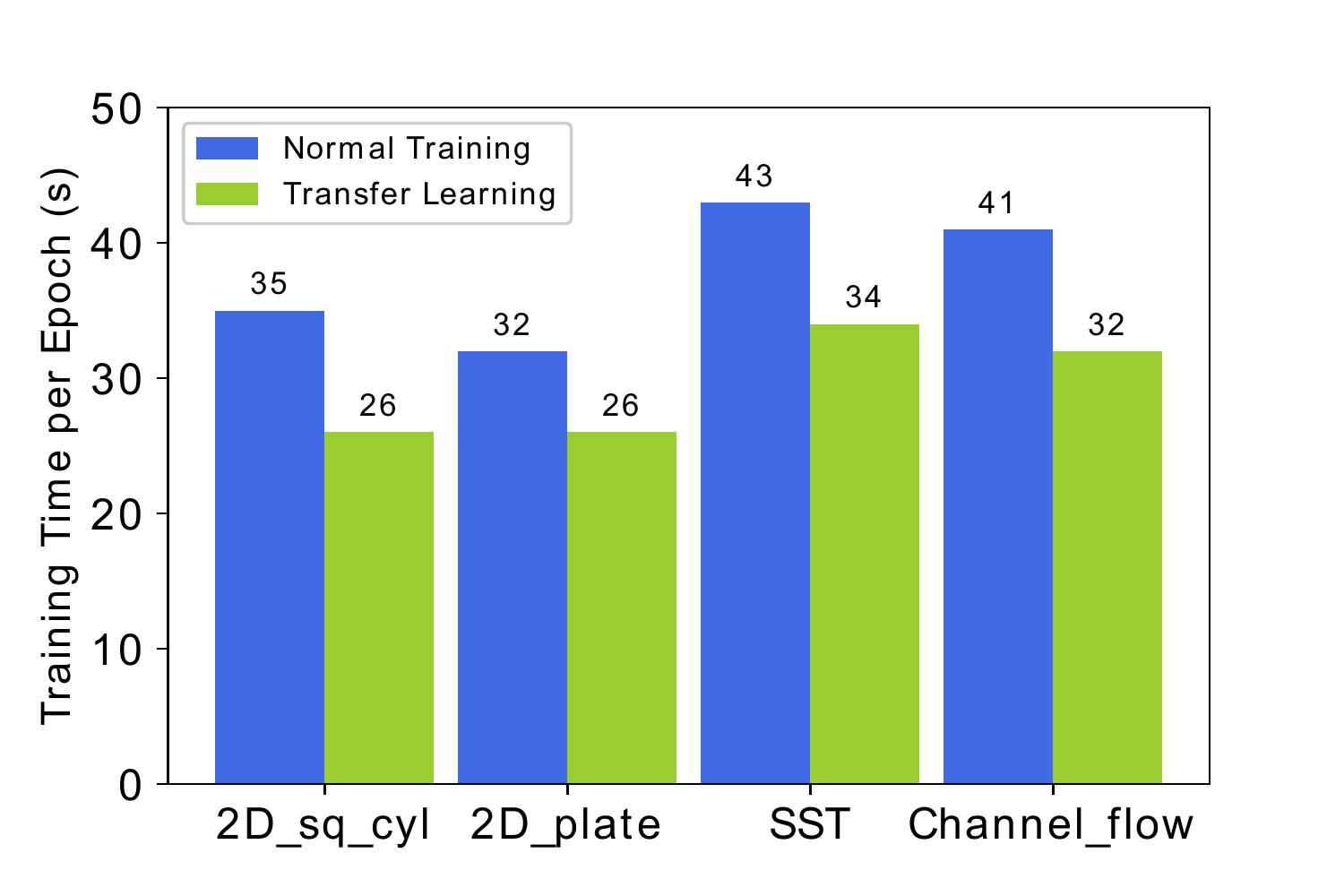}}\hfill
     \subfloat[][]{\includegraphics[width=0.5\linewidth]{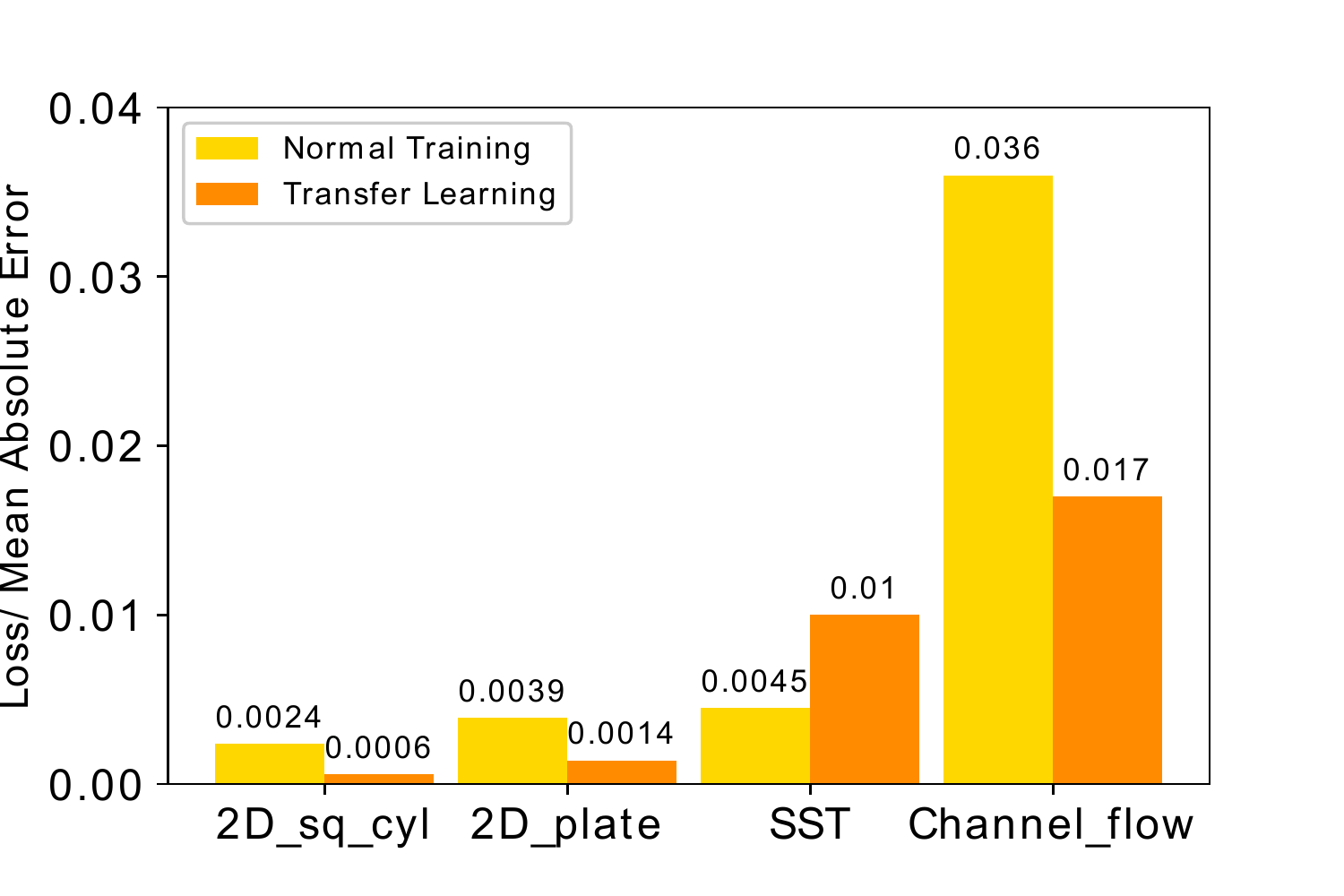}}
      \caption{(a) Comparison of training time per epoch for DL-ROM on different datasets with and without transfer learning. Weights from the training for the 2D cylinder dataset are transferred to the other models before starting training. Notice the decrease in training times when pre-trained weights from the 2d cylinder dataset are directly used while training only the initial and final layer of the DL-ROM model as represented by Fig \ref{fig:architecture}.
      (b) Comparison of Loss/ Mean Absolute Error b/w normal training and training with transfer learning. Using transfer learning we see a considerable speed up in our training times while maintaining comparable performance to conventional training on most datasets.
      }
      \label{fig:transfer_learning}
\end{figure}

\section{Results}

In order to experimentally evaluate our approach, we trained our custom architecture (DL-ROM) on the discussed datasets. Each dataset was split into training and validation set in a ratio of approximately 6:1. The loss function used for training our model was Mean Absolute Error (MAE). The varied input sizes of different datasets were handled in the initial and final layers of the encoder and decoder respectively and the remaining part of the model was kept same for all datasets. By keeping the majority of the network layers the same across different datasets we were able to explore the possibility of implementing transfer learning. In neural networks, transfer learning refers to the prior initialization of network weights by adopting the learned weights of a previously solved similar problem. Previous works such as \citep{Guastoni_2020} have successfully utilized this technique to speed up the training times of ML models. Using an approach similar to aforementioned reference, we were able to train models on different datasets (2D sq. cylinder, 2D plate, Channel Flow, SST) using the weights of a model previously trained on the 2D Cylinder dataset. To make the most use of this feature, we designed the network architecture for DL-ROM in such a way that that apart from the initial and final network layers, the network architecture was independent of the dataset being used. Therefore, while implementing transfer learning, only the initial and final layers of the DL-ROM model were required to be trained from scratch based on the dataset in use. By performing transfer learning we consistently saw a considerable speed up in our training times which is in accordance to the findings of \citep{Guastoni_2020}, while maintaining comparable performance to conventional training on most datasets (Figure \ref{fig:transfer_learning}). \\

As alluded to before, the main objective of our implementation (DL-ROM) was to create the reduced order embeddings of the high order dynamical systems and to use these reduced states to predict future timesteps. To predict the future timesteps, we have used 10 previous frames, each separated by 10 timestpes. This is a hyperparameter and can be varied if needed. Different latent sizes for the reduced order model state were experimented, with size 32 giving a best reconstruction performance while reducing the computational overhead for the reduced state representation.

\begin{figure}
     \centering
     \subfloat[][]{\includegraphics[width=0.5\linewidth]{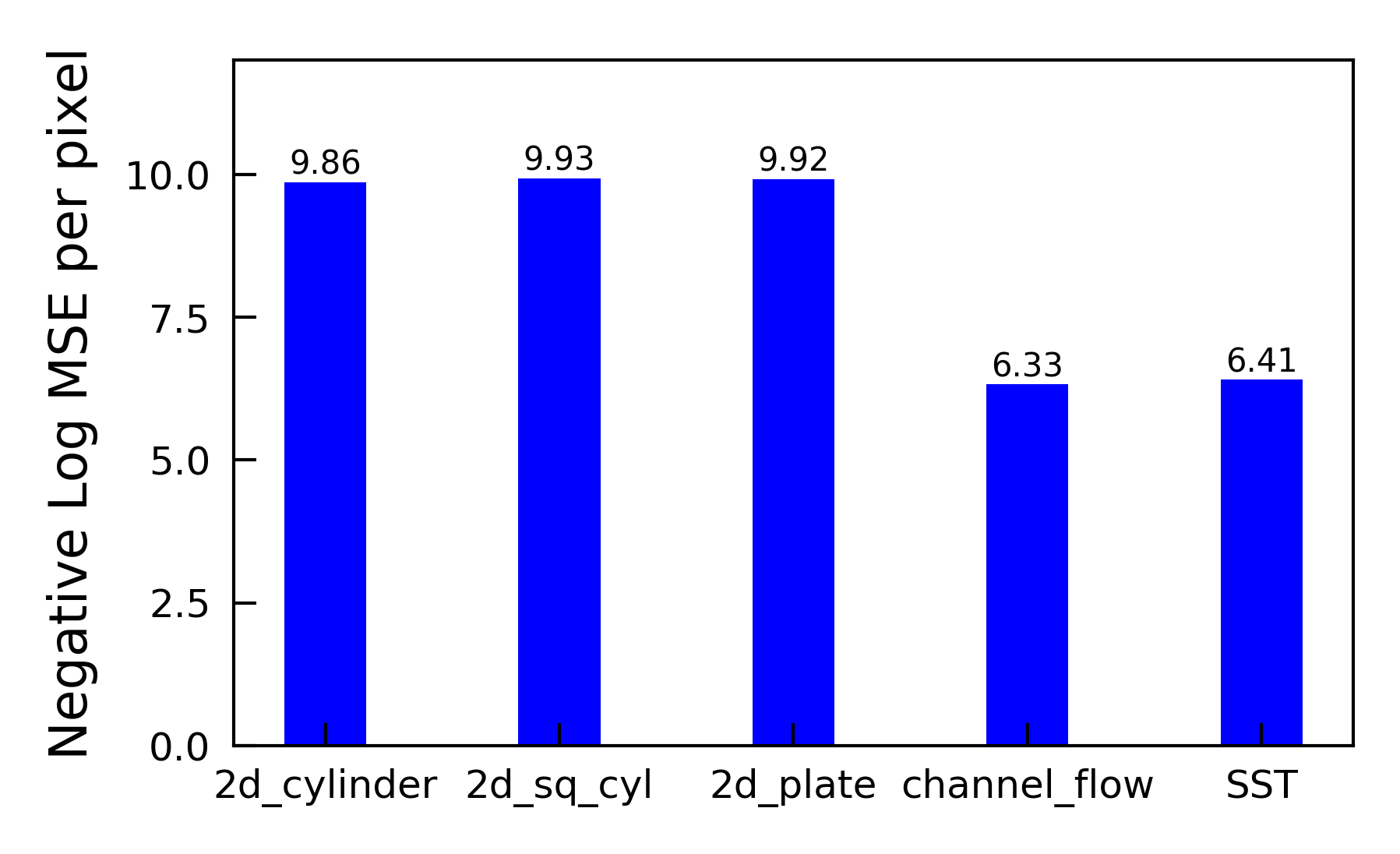}}\hfill
     \subfloat[][]{\includegraphics[width=0.45\linewidth]{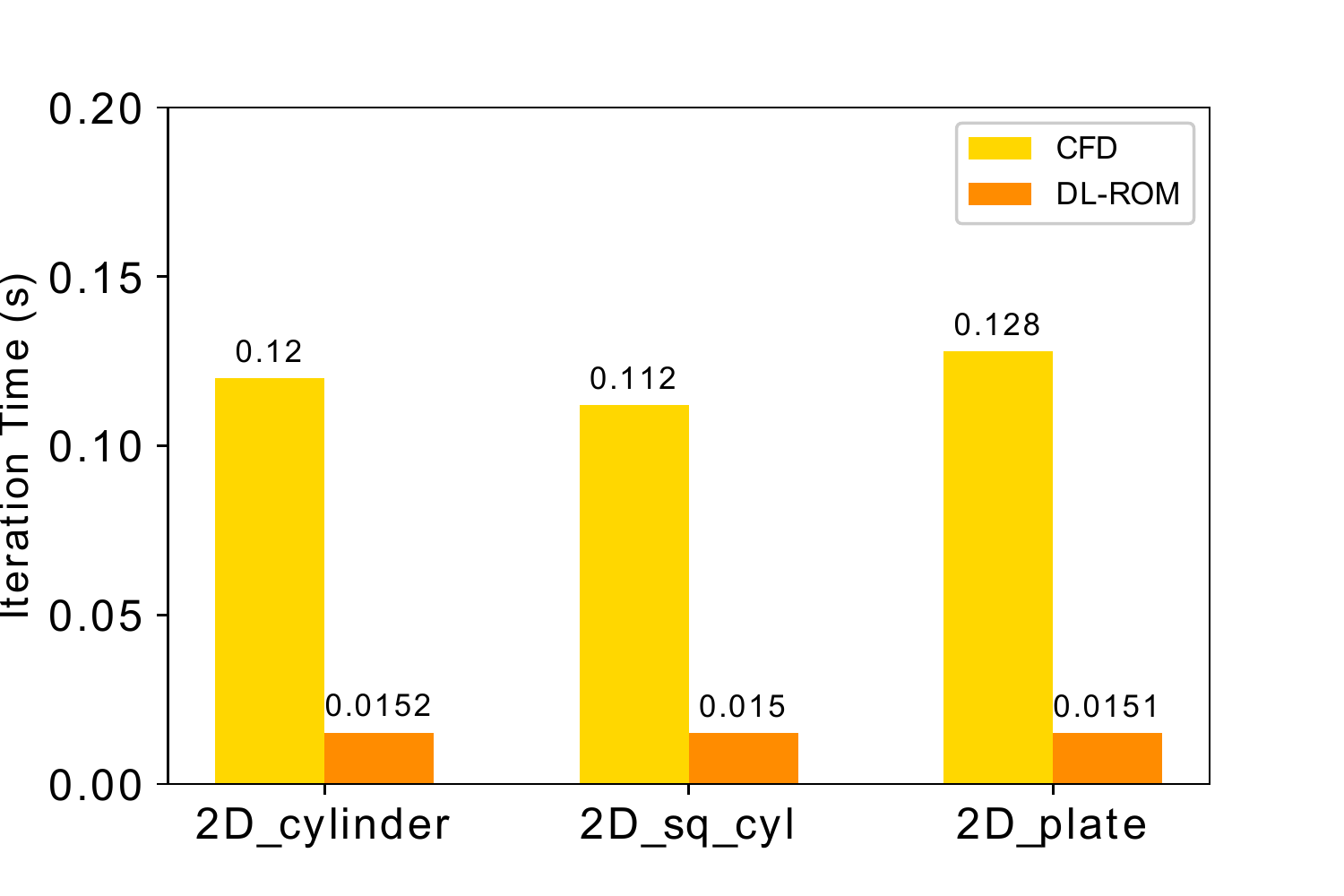}}
      \caption{(a) Average Negative Logarithmic Mean Squared Error per pixel for the 5 studied datasets. Note that higher values of the bar represents better performance.
      (b) Comparing average CPU runtime for one iteration of the simulation. Comparison has been made between CFD (solved on OpenFOAM using the PimpleFOAM solver) and the DL-ROM machine learning model. The DL-ROM outperforms the runtimes of CFD simulations by nearly 2 orders of magnitude
      }
      \label{fig:time}
\end{figure}

Figure \ref{fig:time} shows the average negative logarithm MSE per pixel values for different datasets on their respective validation set. DL-ROM performs well on all the aforementioned datasets. The results table (Figure \ref{fig:domain}) shows the actual label and the predictions of our approach. The error residual image is the subtraction between the labels and predictions. On qualitative evaluation of the results we came across some intersting observations. Owing to the large flow gradients, the region near the obstruction in certain datasets(2D cylinder flow, 2D square cylinder flow, NOAA-SST) is difficult to predict. On the other hand, the complex flow structures such as the von-Karman vortex streets in the wake of these obstructions are accurately predicted for the future timesteps. Additionally, datasets such as 2D Channel flow give sub-optimal reconstruction results owing to the abundance of small flow structures in the flow. The information corresponding to these minute structures is often lost during the process of compression into the reduced state. Next, we created custom CFD datasets (2D cylinder flow, 2D airfoil and 2D flat plate) on OpenFOAM in order to accurately compare the computational runtimes of our model with an iterative CFD solver. Using our approach we were able to train a network that could predict future iterations of the fluid simulation while significantly reducing the iteration time for generating the next timestep. 
From Fig. \ref{fig:time} it is apparent that DL-ROM outperforms the computational runtimes of CFD simulations by nearly 2 orders of magnitude across all three datasets. Also, from Fig. \ref{fig:domain} and \ref{fig:time} we can decipher that the MSE b/w the predictions and the ground truth is very small, particulary for 2D cylinder, 2D airfoil and 2D plate datasets. Finally, from Fig.\ref{fig:domain},\ref{fig:time} we can see that DL-ROM performs well on the looped simulation prediction task. From the lineplots we can clearly seen a gradual decrease in the negative log MSE values over the 20 iterations. This decrease can be attributed to the accumulation of errors over time, but importantly this decrease is very gradual and doesn't result in a sudden departure of the prediction from the ground truth.

\section{Conclusions}

Through this work we present a novel approach to use deep neural networks to create learned reduced order embeddings for representing the dynamics of the systems. Using the learned reduced state we are able to predict the flow state at future time steps given the reduced state of the previous time steps in a highly computationally efficient manner. Our implementation also includes a decoder architecture that reconstructs the reduced state to a highly accurate approximation of the higher order state, by minimizing the reconstruction error with the ground truth simulation of the future time step. To achieve this functionality we combined state of the art advancements made in the realm of neural networks such as 3D-Autoencoders, 3D U-Nets etc. while eliminating the downsides brought about by the use of LSTMs in previous implementations and applied it to the field of ROMs and temporal evolution of fluid simulations. Additionally, deep learning techniques such as depth wise separable convolutions and transfer learning are used to create a novel neural network architecture (DL-ROM). This architecture takes in fluid flow snapshots from 10 previous timesteps, creates a reduced order model and predicts the full order flow state at a future timestep. To evaluate the effectiveness and performance of the network, we test its reconstruction performance on 5 different datasets namely, 2D cylinder, 2D square cylinder, 2D plate, 2D channel flow and the Sea Surface Temperature(SST) datasets. Our implementation yields excellent reconstruction results and is able to accurately predict the future flow with some minor flow differences mostly around regions with high value of gradients. Also owing to our use of a learned reduced order state our network is able to predict future simulation timesteps in significantly reduced computational runtimes when compared to CFD solvers. We observe nearly a 2 orders of magnitude reduction in computational runtimes when compared to CFD solvers (OpenFOAM). Thus, using autoencoder based 3D-UNets to generate reduced order models and subsequently using this reduced state to predict future timesteps presents a novel and effective solution for reducing computational runtimes of CFD simulations. Finally, we also deploy our network to make looped prediction simulations in which the simulation evolves using the predictions of the previous timesteps without supervision from the ground truth values. DL-ROM yields great reconstruction results for this simulation by maintaining low values of MSE over a span of 20 iterations. Hence, we are able to successfully demonstrate that by using a deep neural networks such as ours, we can augment CFD solvers to accelerate the computationally expensive process of iteratively solving Navier-Stokes equations, which can drastically improve the turnaround time for CFD simulations. Thus, we can solve the initial iterations of CFD simulations using conventional, computationally expensive iterative solvers and subsequently hand off the evaluation of future iterations to our deep learning model. DL-ROM would be able to continue the subsequent evaluation of the simulation at a fraction of the computational cost of the iterative solver, while maintaining an acceptable level of error tolerance from the ground truth. Alternatively, evaluations from DL-ROM can also be interleaved between sparse evaluations by iterative CFD solvers. These interleaved CFD evaluations would act to correctively recalibrate the future evaluations so as to prevent the large accumulation of errors over time.


\section{Acknowledgments}
This work is supported by the start-up fund provided by CMU Mechanical Engineering.

\bibliographystyle{unsrtnat}
\bibliography{main}

\end{document}